\documentclass[sigconf]{acmart}
\usepackage{graphicx}
\usepackage{booktabs}
\usepackage{multirow}

\AtBeginDocument{%
  }

\begin{document}

\setcopyright{acmcopyright}
\acmYear{2025}
\copyrightyear{2025}
\setcopyright{acmlicensed}
\acmConference[FSE Companion '25]{Companion Proceedings of the 33rd ACM International Conference on the Foundations of Software Engineering}{June 23--28, 2025}{Trondheim, Norway}
\acmBooktitle{Companion Proceedings of the 33rd ACM International Conference on the Foundations of Software Engineering (FSE Companion '25), June 23--28, 2025, Trondheim, Norway}
\acmDOI{10.1145/3696630.3728519}
\acmISBN{979-8-4007-1276-0/25/06}

\title{From Few-Label to Zero-Label: An Approach for Cross-System Log-Based Anomaly Detection with Meta-Learning}


\author{Xinlong Zhao}
\orcid{0009-0001-8230-1039}
\affiliation{
  \institution{Peking University}
  \city{Beijing}
  \country{China}
}
\email{xinlongzhao1126@gmail.com}

\author{Tong Jia}
\authornotemark[1]
\orcid{0000-0002-5946-9829}
\affiliation{
  \institution{Peking University}
  \city{Beijing}
  \country{China}
}
\email{jia.tong@pku.edu.cn}

\author{Minghua He}
\orcid{0000-0003-4439-9810}
\affiliation{
  \institution{Peking University}
  \city{Beijing}
  \country{China}
}
\email{hemh2120@stu.pku.edu.cn}

\author{Yihan Wu}
\orcid{0009-0006-9969-4795}
\affiliation{
  \institution{National Computer Network Emergency Response Technical Team/Coordination Center of China}
  \city{Beijing}
  \country{China}
}
\email{yhwu1988@163.com}

\author{Ying Li}
\orcid{0000-0001-9667-2423}
\affiliation{
  \institution{Peking University}
  \city{Beijing}
  \country{China}
}
\email{li.ying@pku.edu.cn}

\author{Gang Huang}
\orcid{0000-0002-4686-3181}
\affiliation{
  \institution{Peking University}
  \city{Beijing}
  \country{China}
}
\email{hg@pku.edu.cn}
\renewcommand{\shortauthors}{Xinlong Zhao et al.}

\begin{abstract}
Log anomaly detection plays a critical role in ensuring the stability and reliability of software systems. However, existing approaches rely on large amounts of labeled log data, which poses significant challenges in real-world applications. To address this issue, cross-system transfer has been identified as a key research direction. State-of-the-art cross-system approaches achieve promising performance with only a few labels from the target system. However, their reliance on labeled target logs makes them susceptible to the cold-start problem when labeled logs are insufficient. To overcome this limitation, we explore a novel yet underexplored setting: zero-label cross-system log anomaly detection, where the target system logs are entirely unlabeled. To this end, we propose FreeLog, a system-agnostic representation meta-learning method that eliminates the need for labeled target system logs, enabling cross-system log anomaly detection under zero-label conditions. Experimental results on three public log datasets demonstrate that FreeLog achieves performance comparable to state-of-the-art methods that rely on a small amount of labeled data from the target system.
\end{abstract}

\begin{CCSXML}
<ccs2012>
   <concept>
       <concept_id>10011007.10011006.10011073</concept_id>
       <concept_desc>Software and its engineering~Software maintenance tools</concept_desc>
       <concept_significance>500</concept_significance>
       </concept>
 </ccs2012>
\end{CCSXML}

\ccsdesc[500]{Software and its engineering~Software maintenance tools}

\keywords{Meta-learning, Unsupervised domain adaptation, Anomaly detection, System logs}

\received{17 January 2025}
\received[accepted]{13 March 2025}

\maketitle

\begin{figure*}[h!]
\centering
\includegraphics[width=6.6in]{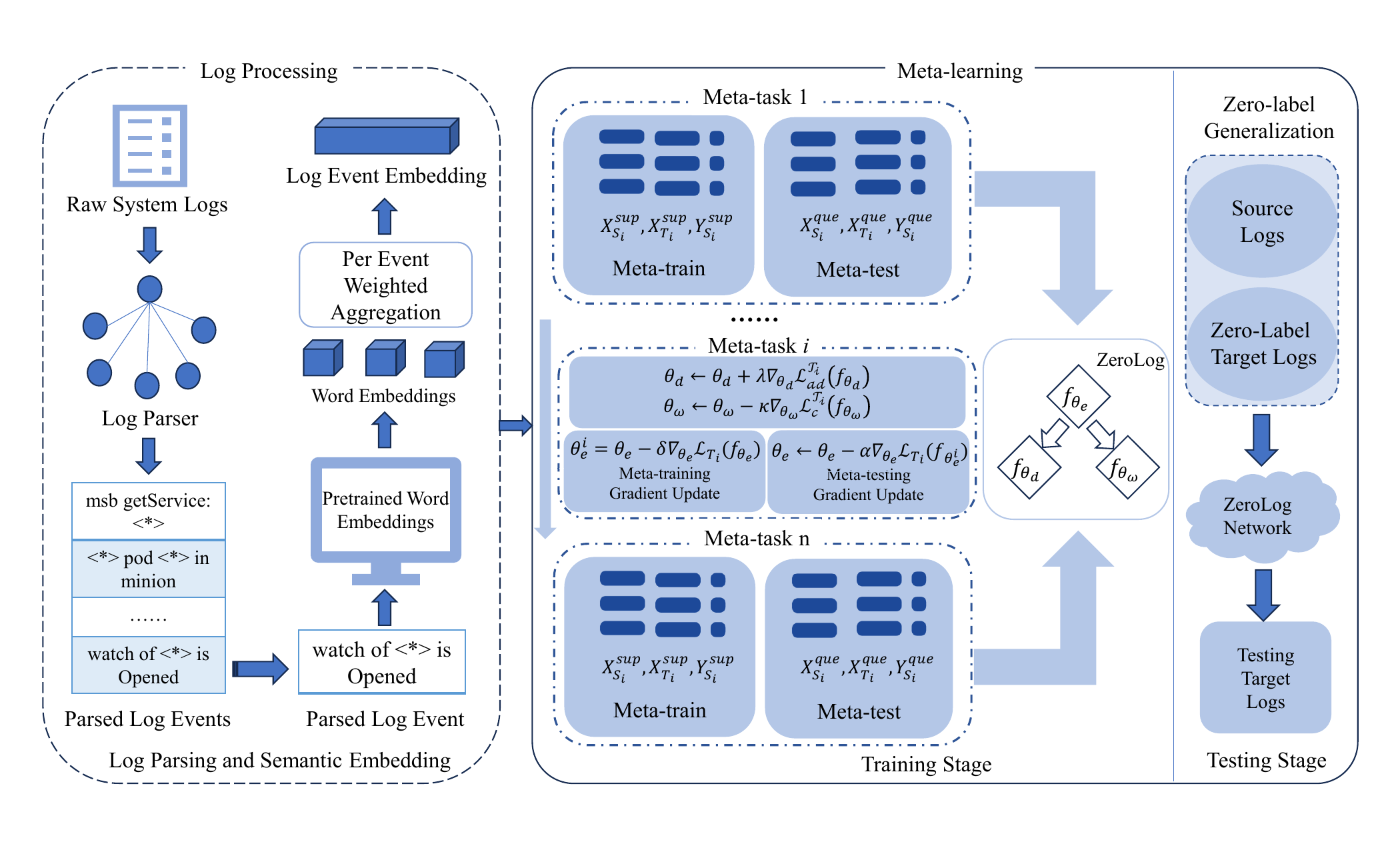}
\vspace{-0.7cm}
\caption{The overview of FreeLog.}
\label{fig_2}
\end{figure*}
\vspace{-0.6cm}

\section{Introduction}
As the scale and complexity of software systems continue to grow, the frequency of system failures increases. Logs, which record critical events and states during system operation, serve as a vital source of information for anomaly detection~\cite{eagerlog, midlog, famos, yang2025enhancing, logcae, llmelog, afalog, aclog, 10.1145/3133956.3134015, 10.1145/3338906.3338931}. Log-based anomaly detection techniques have broad applications in improving system reliability. Existing log-based anomaly detection models can be categorized into unsupervised and supervised approaches. Unsupervised models~\cite{10.1145/3133956.3134015, 10.1145/3377813.3381371, ijcai2019p658, 9240683} use sequential neural networks to learn the occurrence probabilities of log events within normal event sequences. These models predict subsequent log events and detect anomalies based on the discrepancy between the actual and predicted events. However, their detection performance is limited by the lack of labeled anomalous logs~\cite{9401970}. In contrast, supervised models~\cite{5488459, 8854736, 10.1145/3338906.3338931} build classification models to identify anomalous logs, achieving superior detection performance. However, their effectiveness heavily depends on the availability of a large amount of labeled logs.

In real-world software systems, obtaining precise labels is challenging due to the sheer volume of logs and the rarity of anomalous logs, which are often buried within a large number of normal logs~\cite{10.1145/3534678.3539106}. As technology advances, reducing reliance on labeled logs has become a primary research focus, with two main strategies emerging in the literature. The first strategy employs active learning techniques~\cite{10248257, 10.1145/3534678.3539106}, where human annotations are incorporated in real-time to acquire log labels. However, the real-time annotation process is complex and requires significant human effort. In addition, anomaly detection models relying on real-time annotations face the cold-start problem, where model performance depends heavily on the accumulation of annotations. The second strategy involves cross-system transfer for log anomaly detection~\cite{9251092, 10.1145/3459637.3482209, 10.1145/3597503.3639205}. This approach leverages the rich features of labeled logs from mature systems to build anomaly detection models for new systems, effectively transferring the capabilities of models to target systems with limited labeled data. Using knowledge of source systems, this method has demonstrated significant advantages.

Several cross-system log anomaly detection methods, such as LogTransfer~\cite{9251092}, LogTAD~\cite{10.1145/3459637.3482209} and MetaLog~\cite{10.1145/3597503.3639205}, have been proposed. These methods rely on abundant labeled logs in the source systems and a small number of labeled logs in target systems to build anomaly detection models. LogTransfer and LogTAD utilize simple transfer learning models in which the source and target systems share portions of the neural network architecture. However, studies have shown that transfer learning methods only guarantee performance under specific assumptions and may face significant challenges when there are substantial distributional differences between the source and target domains~\cite{10.1007/s10994-009-5152-4, 7542175, ijcai2022p496}. In contrast, MetaLog employs meta-learning to enhance generalization capabilities. Meta-learning involves outer-loop optimization, enabling MetaLog to address more comprehensive meta-representations~\cite{pmlr-v70-finn17a, 9428530}. Research has demonstrated that meta-learning achieves comparable generalization results with less data compared to transfer learning, especially when target data availability is limited~\cite{NEURIPS2020_cfee3986, gu-etal-2018-meta, 10714339}. Consequently, the state-of-the-art cross-system log anomaly detection method, MetaLog, achieves excellent performance with only a small amount of labeled data from the target system. However, MetaLog's performance remains reliant on labeled target logs, limiting its applicability in real-world scenarios with no labeled data and failing to fully address the cold-start problem. 

To tackle this limitation, we investigate a novel yet valuable setting: zero-label cross-system log anomaly detection, where the target system logs are entirely unlabeled. To this end, we propose a new method, FreeLog, which is the first to eliminate the reliance on labeled target system logs. Specifically, to achieve zero-label generalization, we design a system-agnostic representation meta-learning method that leverages both the anomalous classification features of labeled logs from the source domain and domain-invariant features between the source and target domains. We leverage unsupervised domain adaptation techniques to perform adversarial training between the source domain (labeled) and the target domain (unlabeled), aiming to learn system-agnostic general feature representations. By employing meta-learning, the learned system-agnostic general feature representations are further generalized to the target system, enabling zero-label cross-system log anomaly detection. We evaluate the performance of FreeLog on three public log datasets from different systems (HDFS, BGL and OpenStack). Experimental results show that under zero-label conditions, FreeLog achieves an F1-score exceeding 80\%, demonstrating performance comparable to state-of-the-art cross-system log anomaly detection methods trained with partially labeled logs.
\vspace{-0.5cm}
\section{Method}
\subsection{Overview}
Under zero-label conditions, the feature representations of the target system are entirely unknown, making zero-label cross-system log anomaly detection particularly challenging. To overcome these challenges, we propose a novel method, FreeLog, designed to learn and transfer system-agnostic general feature representations for zero-label cross-system log anomaly detection. In particular, FreeLog involves three key processes: Log Embedding, System-Agnostic Representation Meta-Learning and FreeLog Network.

FreeLog begins by employing the classic log parsing technique Drain~\cite{8029742} to process raw logs and extract log events. Compared to traditional index-based methods, semantic embeddings have been shown to provide more informative representations. Therefore, we generate semantic embeddings for each log event using methods consistent with prior studies~\cite{9251092, 10.1145/3459637.3482209, 9401970, 10.1145/3338906.3338931}. To account for the cross-system nature of log anomaly detection, we adopt the global consistency semantic embedding approach inspired by MetaLog~\cite{10.1145/3597503.3639205}, which ensures consistency in event representations by constructing semantic embedding vectors for log events within a shared global space. Finally, these log event embeddings are fed into the FreeLog network for anomaly detection.

To achieve cross-system log anomaly detection without utilizing any labeled target system logs, we introduce an additional domain classifier, which classifies source domain data as $0$ and target domain data as $1$, facilitating domain classification. The domain classifier maximizes the distinction between source and target domain features while leveraging adversarial training to achieve domain alignment. This process facilitates the network in acquiring the necessary generalization capability between source system logs and target system logs. Overall, we learn anomaly classification features from labeled logs in the source domain, enabling our model to effectively distinguish between normal and anomalous logs. Additionally, we learn domain classification features from unlabeled logs in both the source and target domains, employing adversarial training to capture domain-invariant features between the two domains. Based on these two types of features, FreeLog achieves robust cross-system performance under zero-label conditions. The complete workflow of FreeLog is illustrated in Figure \hyperref[fig_2]{1}.
\vspace{-0.5cm}
\subsection{System-Agnostic Representation Meta-Learning}
\label{sec3.2}
There are two challenges in zero-label conditions. First, given the inaccessibility of log labels from the target system, how can we effectively learn feature representations of the target system from the source system? Second, considering the significant differences across software systems, how can the learned feature representations be generalized to the target system?

To overcome these challenges, we propose FreeLog. Specifically, to address the first challenge, we use unsupervised domain adaptation techniques to perform adversarial training between the source and target domains, enabling the learning of system-agnostic feature representations. UDA addresses the challenge of limited model generalization caused by distribution discrepancies between the source and target domains. UDA methods based on adversarial learning extract domain-invariant features, achieving alignment between the source and target domain distributions. These methods typically comprise a feature extractor and a domain classifier. The feature extractor is optimized to generate features indistinguishable by the domain classifier, thereby mitigating domain-specific biases while retaining task-relevant information. This adversarial optimization fosters robust feature representations that generalize effectively across domains. To tackle the second challenge, we apply meta-learning techniques to transfer the learned system-agnostic representations to the target system. Meta-learning enhances a model's adaptability to new tasks by extracting shared task-relevant features and optimizing learning strategies during multi-task training. By leveraging parameter sharing, meta-learning captures commonalities across tasks while suppressing domain-specific variations. Furthermore, it optimizes the model’s initialization to enable rapid convergence on unseen tasks with minimal gradient updates. During inference, the model efficiently adapts to task-specific requirements through fast fine-tuning or direct utilization of shared representations, ensuring strong generalization and adaptability.

The setup of FreeLog involves two domains: the source domain \( D_S \) and the target domain \( D_T \). Formally, \( X_S \) and \( X_T \) represent the data sampled from \( D_S \) and \( D_T \), respectively, while \( Y_S \) denotes the label matrix for \( X_S \). In meta-learning, the smallest unit for gradient updates is the meta-task. A meta-task consists of two phases: the meta-training phase and the meta-testing phase. We construct a meta-task \( T_i = \{ M_i^{sup}, M_i^{que} \} \), where \( M_i^{sup} = \{ X_{S_i}^{sup}, X_{T_i}^{sup}, Y_{S_i}^{sup} \} \) is used for meta-training, and \( M_i^{que} = \{ X_{S_i}^{que}, X_{T_i}^{que}, Y_{S_i}^{que} \} \) is used for meta-testing. FreeLog consists of three main modules: feature extractor \( f_{\theta_e} \), anomaly classifier \( f_{\theta_\omega} \) and domain classifier \( f_{\theta_d} \). For each meta-task, the anomaly classifier attempts to find a local optimal solution for the classification problem, while both the anomaly classifier and domain classifier iteratively update across tasks to find the global optimal solution. In the following, we will elaborate on our method, which can be divided into three key components.

\textit{Training the Domain Classifier and Anomaly Classifier.} Given the feature extractor \( f_{\theta_e} \), we train the domain classifier \( f_{\theta_d} \) on the meta-task to distinguish between the source and target domain features. The optimization problem is: $\max_{\theta_d} \sum_{T_i \sim p(T)} \mathcal{L}_{ad}^{T_i}(M_i^{sup}; f_{\theta_d})$. Adversarial training encourages the feature extractor to learn domain-invariant features that are shared between the source and target domains. Next, we train the anomaly classifier \( f_{\theta_\omega} \) to distinguish between normal and anomalous logs. The optimization problem is: $\min_{\theta_\omega} \sum_{T_i \sim p(T)} \mathcal{L}_c^{T_i}(M_i^{sup}; f_{\theta_\omega})$. The objective is to learn discriminative features for classifying normal and anomalous logs, thereby achieving robust anomaly detection performance.

\begin{table*}[h!]
\centering
\setlength{\belowcaptionskip}{0.3cm}
\caption{Zero-label generalization experiments across different domains.}
\label{tab_all}
\resizebox{\textwidth}{!}{%
\begin{tabular}{@{}lcccccccccccc@{}}
\toprule
\multicolumn{1}{c}{\multirow{2}{*}{\textbf{Method}}} & \multicolumn{3}{c}{\textbf{HDFS to BGL}} & \multicolumn{3}{c}{\textbf{BGL to HDFS}} & \multicolumn{3}{c}{\textbf{OpenStack to HDFS}} & \multicolumn{3}{c}{\textbf{OpenStack to BGL}} \\
\cmidrule(lr){2-4} \cmidrule(lr){5-7} \cmidrule(lr){8-10} \cmidrule(lr){11-13}
 & \textbf{Precision} & \textbf{Recall} & \textbf{F1-score} & \textbf{Precision} & \textbf{Recall} & \textbf{F1-score} & \textbf{Precision} & \textbf{Recall} & \textbf{F1-score} & \textbf{Precision} & \textbf{Recall} & \textbf{F1-score} \\
\midrule
(a1) PLELog~\cite{9401970} & 82.10 & 67.42 & 74.04 & 65.86 & 71.11 & 68.38 & 65.86 & 71.11 & 68.38 & 82.10 & 67.42 & 74.04 \\
(a2) LogRobust~\cite{10.1145/3338906.3338931} & 94.60 & 72.95 & 82.38 & 100.00 & 62.30 & 76.77 & 100.00 & 62.30 & 76.77 & 94.60 & 72.95 & 82.38 \\ \hline
(b1) DeepLog~\cite{10.1145/3133956.3134015} & 66.13 & 48.79 & 56.16 & 53.96 & 34.07 & 41.77 & 53.96 & 34.07 & 41.77 & 66.13 & 48.79 & 56.16 \\ \hline
(c1) PLELog~\cite{9401970} & 38.80 & 99.87 & 55.89 & 1.69 & 92.85 & 3.32 & 4.33 & 53.47 & 8.01 & 54.65 & 43.04 & 48.16 \\
(c2) LogRobust~\cite{10.1145/3338906.3338931} & 39.08 & 93.67 & 55.15 & 2.25 & 62.12 & 4.35 & 0.63 & 60.81 & 1.25 & 34.34 & 57.39 & 42.97 \\
(c3) NeuralLog~\cite{9678773} & 57.23 & 52.79 & 54.38 & 33.13 & 58.04 & 42.06 & 3.33 & 42.99 & 6.17 & 14.76 & 81.45 & 24.99 \\ \hline
(d1) LogTAD~\cite{10.1145/3459637.3482209} & 78.01 & 68.51 & 72.95 & 78.80 & 71.22 & 74.82 & 71.89 & 65.51 & 68.55 & 70.72 & 65.51 & 68.02 \\
(d2) LogTransfer~\cite{9251092} & 74.42 & 76.73 & 75.56 & 100.00 & 43.30 & 60.43 & 73.87 & 62.30 & 67.59 & 68.43 & 71.32 & 69.85 \\ \hline
(e1) MetaLog~\cite{10.1145/3597503.3639205} & 96.89 & 89.28 & 92.93 & 89.29 & 74.98 & 81.51 & 96.67 & 62.42 & 75.86 & 99.83 & 70.09 & 82.36 \\ \hline
Ours FreeLog & 83.10 & 89.13 & 86.01 & 82.70 & 78.61 & 80.61 & 75.73 & 85.24 & 80.21 & 74.44 & 80.92 & 77.55 \\
\bottomrule
\end{tabular}%
}
\end{table*}

\textit{Adapting to Meta-Tasks.} To train a feature extractor \( f_{\theta_e} \) that can effectively adapt to the target system's features, we learn to adapt the meta-task \( T_i \). During the meta-training phase, the learner's parameters \( \theta_e \) become \( \theta_e^i \), which can be updated through one or more gradient descent steps: $\theta_e^i = \theta_e - \delta \nabla_{\theta_e} \mathcal{L}_{T_i}(M_i^{sup}; f_{\theta_e})$, where \( \delta \) is the learning rate. In the task \( T_i \), the goal is to achieve effective classification while performing domain adaptation. Therefore, the objective function for the current task can be written as:
\[
\mathcal{L}_{T_i}(f_{\theta_e}) = \gamma \mathcal{L}_c^{T_i}(X_{S_i}^{sup}, Y_{S_i}^{sup}; f_{\theta_\omega}) + \beta \mathcal{L}_{ad}^{T_i}(X_{S_i}^{sup}, X_{T_i}^{sup}; f_{\theta_d}).
\]
The first term represents the classification loss in the source domain with labeled information. The second term is the domain adversarial loss, which encourages the feature extractor \( f_{\theta_e} \) to produce domain-invariant features by aligning the domains through the domain classifier \( f_{\theta_d} \). The hyperparameters \( \beta \) and \( \gamma \) control the trade-off between adaptation and classification performance. This method integrates classification loss and adversarial loss, enabling FreeLog to effectively generalize from the source system to the target system.

\textit{Meta-Optimization of Tasks.} After learning the adaptation parameters \( \theta_e^i \) for each task, we proceed to meta-optimize the feature extractor \( f_{\theta_e} \). The objective is to improve the performance of \( \theta_e^i \) on the query set. The meta-objective function can be expressed as: $\min_{\theta_e} \sum_{T_i \sim p(T)} \mathcal{L}_{T_i}(M_i^{que}; f_{\theta_e^i})$. We perform meta-optimization as follows: $\theta_e \gets \theta_e - \alpha \nabla_{\theta_e} \sum_{T_i \sim p(T)} \mathcal{L}_{T_i}(M_i^{que}; f_{\theta_e^i})$, where \( \alpha \) is the meta-step size. During meta-optimization, we apply domain adversarial training to the query set through the domain classifier, which ensures domain alignment. Therefore, the goal of the meta-optimization process is to learn a general feature extractor that can quickly adapt to new tasks involving both classification and domain adaptation. Overall, the total loss for FreeLog can be expressed as a min-max optimization problem:
\[
\min_{\theta_e} \max_{\theta_d} \sum_{T_i \sim p(T)} \left( \gamma  \mathcal{L}_c^{T_i}(f_{\theta_e}, f_{\theta_\omega}) + \beta  \mathcal{L}_{ad}^{T_i}(f_{\theta_e}, f_{\theta_d}) \right).
\]
This integrated optimization strategy takes into account information from both the source and target systems, providing sufficient cues for the FreeLog network to solve the zero-label cross-system log anomaly detection task.

\section{Experiments}
We conducted experiments on three publicly available log datasets: HDFS~\cite{10.1145/1629575.1629587}, BGL~\cite{4273008} and OpenStack~\cite{10.1145/3133956.3134015}. In the zero-label setting, we selected four cross-system dataset combinations (HDFS to BGL, BGL to HDFS, OpenStack to HDFS and OpenStack to BGL) to validate our method. In these experiments, the labeled logs from the source system and the unlabeled logs from the target system were used to train the FreeLog network. Given the variability in dataset sizes, we adopted distinct proportions for each dataset. To perform zero-label generalization tasks and ensure a fair comparison with our method, we adopted various baseline methods and experimental setups. 

\textbf{Baseline.} Block (a) presents the performance of the semi-supervi-sed method PLELog~\cite{9401970} and the fully supervised baseline LogRobust~\cite{10.1145/3338906.3338931}. When BGL is the target system, the method of block (a) are trained on 30$\%$ of normal log sessions and only 1$\%$ of anomalous log sessions from BGL dataset. When HDFS is the target system, the method of block (a) are trained on 10$\%$ of normal log sessions and 1$\%$ of anomalous log sessions from HDFS dataset. Block (a) evaluates the performance of methods trained solely on the target dataset under a scenario where anomaly labels are scarce. Block (b) presents the unsupervised baseline method DeepLog~\cite{10.1145/3133956.3134015}, which is trained exclusively on normal labels from the target dataset, maintaining consistency with other baseline methods. Block (c) evaluates the performance of PLELog, LogRobust and NeuralLog~\cite{9678773} under the zero-shot generalization setup. These methods are trained solely on subsets of the source datasets (all log sessions of BGL and OpenStack, 30$\%$ of log sessions of HDFS) and are directly tested on subsets of the target datasets. Block (d) highlights two transfer learning baseline methods, LogTAD~\cite{10.1145/3459637.3482209} and LogTransfer~\cite{9251092}. In HDFS to BGL generalization, the method of block (d) are trained on 30$\%$ of normal log sessions and 1$\%$ of anomalous log sessions from BGL dataset and 30$\%$ of log sessions from HDFS dataset. In BGL to HDFS generalization, block (d) presents results for these methods trained on 10$\%$ of normal log sessions and 1$\%$ of anomalous log sessions from HDFS dataset and all log sessions from BGL dataset. Similarly, in OpenStack to HDFS generalization, block (d) reports results for methods trained on 10$\%$ of normal log sessions and 1$\%$ of anomalous log sessions from HDFS dataset and all log sessions from OpenStack dataset. In OpenStack to BGL generalization, block (d) follows the same setup for the BGL and OpenStack dataset. Finally, block (e) showcases a baseline method based on meta-learning, MetaLog~\cite{10.1145/3597503.3639205}, which shares the same source and target data configurations as LogTAD and LogTransfer. We selected precision, recall, and F1-score as evaluation metrics. 

According to table \hyperref[tab_all]{1}, FreeLog demonstrates significant advantages over existing methods in various settings, including direct zero-shot scenarios (Block (c)) and transfer learning methods (Block (d)). Furthermore, FreeLog even outperforms (semi-) supervised methods directly trained on the target log data (Block (a)) in many cases, while substantially surpassing unsupervised methods (Block (b)). Finally, FreeLog achieves comparable performance to recent methods based on meta-learning (Block (e)). Notably, in contrast to MetaLog, FreeLog achieves such superior performance without leveraging target log labels for training.  
\vspace{-0.5cm}
\section{Conclusion}
In this paper, we study a previously unexplored yet highly valuable setting: zero-label generalizable cross-system log anomaly detection, and propose a novel method FreeLog. To achieve zero-label generalization, we design a system-agnostic representation meta-learning method that effectively leverages anomaly classification features from labeled logs in the source domain, as well as domain-invariant features between the source and target domains. In the zero-label generalization scenario, FreeLog achieves an F1-score exceeding 80$\%$, and it performs comparably to the latest cross-system log anomaly detection methods that use labeled logs for training, demonstrating its potential for generalized anomaly detection across diverse software systems. In future research, we will further test the zero-label generalization performance of FreeLog on more types of log datasets and explore its practical applicability.

% \begin{acks}
% \end{acks}

\bibliographystyle{ACM-Reference-Format}
\bibliography{sample-base}


\end{document}